\documentstyle[prd,aps,eqsecnum]{revtex}
\begin{document}
\draft
\preprint{WUGRAV 94-5}
\title{Testing scalar-tensor gravity with gravitational-wave
observations of inspiralling compact binaries}
\author{Clifford M. Will}
\address{McDonnell Center for the Space Sciences,
Department of Physics\\
Washington University, St. Louis, Missouri 63130}
\twocolumn[
\maketitle
\widetext
\newpage
\begin{abstract}
Observations of gravitational waves from inspiralling compact
binaries using laser-interferometric detectors can provide accurate
measures of parameters of the source.  They can also constrain
alternative gravitation theories.  We analyse inspiralling compact
binaries in the context of the scalar-tensor theory of Jordan,
Fierz, Brans and Dicke, focussing on the effect on the inspiral
of energy lost to dipole gravitational radiation, whose source is
the gravitational self-binding energy of the inspiralling bodies.
Using a matched-filter analysis we obtain a bound on the
coupling constant $\omega_{\rm BD}$ of Brans-Dicke theory.
For a neutron-star/black-hole binary, we find that the bound
could exceed the current bound of $\omega_{\rm BD}>500$ from
solar-system experiments, for sufficiently low-mass systems.
For a $0.7 M_\odot$ neutron star and a $3 M_\odot$ black hole
we find that a bound $\omega_{\rm BD} \approx 2000$ is achievable.
The bound decreases with increasing black-hole mass.
For binaries consisting of two neutron stars, the bound
is less than 500 unless the stars' masses differ by more than
about $0.5 M_\odot$.  For two black holes, the behavior of the
inspiralling binary is observationally indistinguishable from
its behavior in general relativity.  These bounds assume reasonable
neutron-star equations of state and a detector signal-to-noise
ratio of 10.
\end{abstract}

\pacs{PACS numbers:  04.30.-w,04.80.Cc\\
Preprint No. WUGRAV 94-6}
]

\narrowtext

\section{Introduction and Summary}

The regular detection of gravitational radiation from astrophysical
sources by large-scale laser-interferometer systems as the US LIGO
or the French-Italian VIRGO projects will usher in a new era of
gravitational-wave astronomy \cite{ligo}.  One of the most promising
sources for detection by laser-interferometric systems is the
inspiralling compact binary, a binary system of neutron stars
or black holes whose orbit is decaying toward a final coalescence
under the dissipative influence of gravitational radiation reaction.
For much of the late-time evolution of such systems, the gravitational
waveform signal is accurately calculable \cite{lincoln}, given by
a ``chirp'' signal, increasing in amplitude, and sweeping in frequency
through the detectors' typical sensitive bandwidth between 10~Hz and
1000~Hz.  Astrophysical estimates of the rate of such inspiral events
are promising: for the advanced version of LIGO, capable of detecting
the inspiral waveform to distances of hundreds of Mpc, the estimated
rate is 3 per year, and could be as large as 100 per year \cite{clark}.

In addition to simple detection of the waves, it will be possible to
determine important parameters of the inspiralling systems, such as
the masses and spins of the bodies \cite{jugger}.  This is made
possible by the technique of matched filtering of theoretical
waveform templates, which depend on the system parameters,
against the broad-band detector output \cite{filters}.
The method exploits the fact that, depending on the source,
between 500 and 16,000 cycles of the waves may be observable in the
sensitive bandwidth, and so the matching of a template to the signal
will be extremely sensitive to the evolution of the gravitational-wave
frequency with time.  That evolution depends, of course, on
gravitational radiation reaction, which depends on the parameters
of the system.  Very accurate determinations of the masses of the
components should be possible, while less accurate estimates of spins
and other parameters may be feasible \cite{finn,cutler}.

It is said that the first detection of gravitational radiation will
also constitute a verification of general relativity, since that is
the basic theory used in all calculations of gravitational radiation
from such systems.  It would be useful, however, to quantify that
statement,  by assessing how accurately such observations could
actually constrain or bound alternative theories of gravity.
This is not a straighforward question to answer with generality.
Whereas in the slow-motion, weak-field, non-radiative limit
appropriate to solar-system dynamics, most alternative metric
theories of gravity can be encompassed by one simple framework,
known as the parametrized post-Newtonian (PPN) formalism
(see \cite{tegp} for review and references), no correspondingly
simple framework exists for describing radiative systems,
or systems containing strong-internal-field, compact objects,
such as neutron stars or black holes.  On the other hand,
for one simple, but popular class of alternatives,
the scalar-tensor theory of Fierz, Jordan, Brans and
Dicke \cite{jordan,brans}, and some of its generalizations,
the full details have been worked out.  In this paper,
we shall explore the extent to which observations of
gravitational waves from inspiralling compact binaries
could usefully constrain scalar-tensor gravity.

For simplicity, we focus on the version of scalar-tensor
gravity known for short as the Brans-Dicke (BD) theory.
That theory augments general relativity (GR) by the addition
of a scalar gravitational field that couples universally
to matter (hence the theory, like GR, is a metric theory,
satisfying all fundamental equivalence principle
tests \cite{equiv}) and determines the gravitational coupling
strength $G$ via $G \propto \phi^{-1}$.  The relative importance
of the scalar field is parametrized by a coupling constant
$\omega_{\rm BD}$ (in generalized scalar-tensor theories,
$\omega_{\rm BD}$ can itself be a function of the scalar field).
Roughly speaking, in the limit of large $\omega_{\rm BD}$,
the relative difference between effects in GR and effects
in BD is $O(1/\omega_{\rm BD})$.   As $\omega_{\rm BD} \to \infty$,
BD tends smoothly toward GR.  The best current empirical bound
is $\omega_{\rm BD} > 500$, from solar-system measurements of
the Shapiro time delay and the deflection of radio waves
by the Sun \cite{shapiro}.

For systems involving gravitational radiation and compact objects,
BD introduces three effects \cite{eardley,tegpwaves,zaglauer}:

(i) Modifications to the effective masses of the bodies.  These
modifications depend on the internal structure of the bodies,
as parametrized by ``sensitivities'' $s_A$, which roughly measure
the gravitational binding energy per unit mass.  These effects
violate the Strong Equivalence Principle \cite{tegpsep},
in that the motion of such bodies now depends on their internal
structure (apart from tidal interactions).  For neutron stars,
$s\approx 0.1-0.2$, and for black holes $s\equiv 0.5$.

(ii) Modifications of quadrupole gravitational radiation.
BD predicts monopole as well as quadrupole gravitational
radiation, whose combined effect is to modify the effective
GR quadrupole formula for two-body energy loss,
\begin{equation}
{{dE} \over {dt}}=-{8 \over 15} {{\mu^2m^2} \over r^4}
(12v^2-11{\dot r}^2) \,,
\end{equation}
by corrections of $O(1/\omega_{\rm BD})$.   Here, $\mu$ and $m$
are the reduced and total mass respectively; $r$, $v$ and $\dot r$
are the orbital separation, velocity and radial velocity, and
the units are such that $G=c=1$.

(iii) Dipole gravitational radiation.  The center of gravitational
binding energy need not be coincident with the fixed center of
inertial mass, if the two bodies are different, and in BD, the
resulting varying dipole moment is a source of scalar radiation.
Because it is a dipole rather than a quadrupole effect, the dipole
contribution to the energy loss has two fewer time derivatives,
and thus is $O(v^{-2})$ larger than the quadrupole contribution.
It also depends on the difference in sensitivities,
${\cal S} \equiv s_1-s_2$, between the two bodies.  Specifically,
\begin{equation}
\biggl({{dE} \over {dt}}\biggr)_{\rm Dipole}
=-{2 \over 3} {{\mu^2m^2} \over r^4}
\biggl({{\cal S}^2 \over { \omega_{\rm BD}}} \biggr) \,.
\end{equation}
We work here to first order in $1/\omega_{\rm BD}$.

The most important consequence of dipole gravitational radiation
is that it modifies the evolution of the orbital radius and thence
the gravitational-wave frequency $f$, because
${\dot f}/f=-(3/2)\dot r /r=-(3/2)\dot E /|E|$.
In the matched filtering method, any difference between the frequency
evolution of the theoretical template and that of the actual signal
will ultimately cause the two to go out of phase, and the
signal-to-noise ratio will drop.  A rough measure of the accuracy
of the template, then, can be obtained by determining how much
of a change in the template is sufficient to cause a change of
$\approx \pi$ radians in the total accumulated gravitational-wave
phase over the cycles in the detector's sensitive bandwidth.
The accumulated phase is
\begin{equation}
\Phi_{GW} = \int_{t_{\rm in}}^{t_{\rm out}} 2\pi fdt
=\int_{f_{\rm in}}^{f_{\rm out}} 2\pi(f/ \dot f ) df \,, \label{phase}
\end{equation}
where the subscripts in and out denote the values when the signal
enters and leaves the detector's bandwidth.  By demanding that the
change in $\Phi_{GW}$ caused by the dipole term be smaller than
$\pi$, one obtains the bound
\begin{equation}
{{\cal S}^2 \over \omega_{\rm BD}} < {{5376 \pi} \over 25}
(\pi {\cal M} f_{\rm in})^{7/3}\eta^{-2/5} \,, \label{bound1}
\end{equation}
where $\eta=\mu/m$, and ${\cal M}=\eta^{3/5} m$ is the ``chirp mass'',
the mass that determines the lowest-order quadrupole effects.  For
LIGO/VIRGO systems, $f_{\rm in}$ is typically chosen to be 30 Hz.
A more accurate estimate obtained using the formalism of matched
filtering (including post-Newtonian effects -- see below) weakens
this bound by about a factor of 1.3, assuming a signal-to-noise
ratio of 10.  The resulting bound can be fit by the analytic formula
\begin{equation}
{{\cal S}^2 \over \omega_{\rm BD}} < 1.46 \times 10^{-5}
\biggl( {{\cal M} \over {M_\odot}} \biggr) ^{7/3}\eta^{-2/5}
\biggl( {10 \over {S/N}} \biggr)  \,, \label{fit}
\end{equation}

Whether this bound provides a useful constraint on the theory depends
on the system in question.

(i) {\it Neutron star and black hole}.  Since $s_{BH} =0.5$, and
$s_{NS} \le 0.2$, ${\cal S} \ge 0.3$.  The resulting bound on
$\omega_{\rm BD}$ is given, from Eq. (\ref{fit}), by
\begin{equation}
\omega_{\rm BD} > 6156 \,\eta^{2/5}
\biggl( {M_\odot \over {\cal M}} \biggr)^{7/3}
\biggl( {{S/N} \over 10} \biggr)
\biggl( {{\cal S} \over 0.3} \biggr) ^2 \,, \label{bound2}
\end{equation}
where $S/N$ denotes the signal-to-noise ratio of the detected signal.
The resulting bounds on $\omega_{\rm BD}$ are plotted against the
black-hole mass, for various neutron-star masses, in Fig. \ref{fig1}.

(ii) {\it Two neutron stars}.  For neutron stars, $s_{NS}$
varies weakly with mass (see for example Table 3 of \cite{zaglauer}),
so that typically ${\cal S}$ is smaller than 0.05, and for neutron
stars each around $1.4 M_\odot$, is very small indeed.  Thus, the
small value of $\cal M$ in Eq. (\ref{fit}) is compensated by the
smallness of $\cal S$, and the resulting bound on $\omega_{\rm BD}$
is generally weaker than solar-system results unless the difference
in mass between the two neutron stars exceeds about $0.5 M_\odot$.
For the extreme case of $0.7 M_\odot$ and $1.4 M_\odot$ neutron
stars, the bound could be as large as 1100.  The inferred bounds
are sensitive to the assumed equation of state for neutron-star
matter.  For a particular assumption about the dependence of
$\cal S$ on mass, Fig. \ref{fig2} shows the bounds that could
be achieved, assuming a signal-to-noise ratio of 10.

(iii) {\it Two black holes}.  Because $s_{BH} \equiv 0.5$, ${\cal S}
\equiv 0$, and there is no dipole radiation at all (see
\cite{zaglauer} for discussion).  In fact the evolution of
the system and the resulting gravitational radiation are identical
to the general relativistic results, except that the effective
gravitational mass of each black hole is given by Hawking's
``tensor mass'' $m_T$, \cite{hawking} related to the inertial
mass by $m_T=(3+2\omega_{\rm BD})/(4+2\omega_{\rm BD})m$.  Since
the effective gravitational masses are the only parameters
determined from the gravitational-wave signal, no test of BD
is possible from inspiralling two-black-hole systems.

In order to place a bound on $\omega_{\rm BD}$, we must be able to
decide among cases (i), (ii) and (iii).  This requires
that both the chirp mass $\cal M$
and the reduced mass parameter $\eta$ have been measured with
sufficient accuracy that the mass of one of the bodies is known
to be greater than the maximum mass for a neutron star, and
thus is a black hole, and that the mass of the other is known
to be less than the maximum mass, and is thus likely
(though not certain) to be a neutron star, or that the two masses,
if both less than the maximum neutron-star mass, are sufficiently
different to provide an interesting bound.  We have extended
the matched filtering analysis to include post-Newtonian effects
in the evolution of the gravitational-wave frequency, effects
that depend explicitly on the reduced mass parameter $\eta$,
and found that the accuracy in determining $\cal M$ and $\eta$
simultaneously with the dipole radiation effect is sufficient
for this purpose.  The dipole radiation effect varies as
$v^{-2} \approx r/m$ relative to quadrupole radiation,
while post-Newtonian correction terms vary as $m/r$, hence
the two effects are relatively uncorrelated in the matched
filtering (correlation coefficients of order 0.90).

The main result is that for a neutron star of mass typical
of those of well-measured pulsars ($1.3 - 1.5 M_\odot$),
and a relatively light black hole ($3 M_\odot$),
a bound on the Brans-Dicke parameter around two
times the current solar-system bound could be obtained.
For a low-mass system ($0.7 M_\odot$ neutron star, $3 M_\odot$
black hole), a bound around 2000 could be obtained.
For double neutron-star systems of sufficiently different mass,
interesting bounds could also result.

The remainder of this paper provides details.  In Sec. II, we
summarize the relevant BD equations for gravitational radiation
and orbital motion in systems containing compact objects.
Section III derives bounds on the dipole radiation effect
using both the crude estimate of accumulated phase, and a
full matched-filtering analysis.  In Sec. IV we discuss
the possibility of observing BD effects in the gravitational-wave
amplitude, including testing for the existence of a third,
scalar polarization mode in the gravitational waveform.
Section V discusses the results.

\section{Compact objects and gravitational radiation in scalar-tensor
gravity}

The equations of motion and of gravitational radiation for systems
containing compact objects in Brans-Dicke theory were first derived
by Eardley \cite{eardley} and extended by Will and Zaglauer
\cite{tegpwaves,zaglauer}.  Here we present the basic background
and formulas needed for our purpose, taken from \cite{zaglauer}

We work to lowest order in an expansion in powers of
$v^2 \approx m/r$, corresponding to Newtonian order for orbital
motion, and to what in GR would be called quadrupole order
for gravitational radiation.  Yet the equations include contributions
due to the self-gravitational binding energy of the compact objects,
contributions that are formally of post-Newtonian order, but that
could be sizable because of the relativistic nature of neutron stars
or black holes.  These effects are determined by the ``sensitivity''
of the inertial mass of each body A to changes in the local value
of the effective gravitational constant $G$ (caused by changes in
the scalar field).  It is defined by
\begin{equation}
s_A \equiv - \partial (\ln m_A)/\partial (\ln G ) \,.
\end{equation}
In the weak-field limit, it is straightforward to show that $s_A$ is
the negative of the usual gravitational self-energy, given by
$s_A=(1/2)\int_A \rho \rho^\prime |{\bf x}-{\bf x^\prime}|^{-1}
d^3x d^3x^\prime$, where $\rho$ is the mass density.  For neutron
stars, $s$ depends on the equation of state; representative values
are given in Table 3 of \cite{zaglauer}.  For instance, for a
$1.4 M_\odot$ neutron star, $s=0.143$ and $0.125$ for equations
of state O and M respectively.  On the other hand, $s$ can vary
slowly with mass: for equation of state O, for instance,
$s=0.1432$ for $m=1.40 M_\odot$ and $s=0.1440$ for $m=1.44 M_\odot$.
For black holes, a dimensional argument shows that
$m \propto G^{-1/2}$, so $s_{BH} \equiv 0.5$.

\subsection{Two-body orbits}

The equation of motion and Kepler's third law for two-body orbits
with orbital frequency $\omega$ (orbital period $P$) take the form
\begin{equation}
d^2 {\bf x}/dt^2=-{\cal G}m{\bf x} /r^3 \,,\quad
\omega=2\pi/P=({\cal G} m/r^3 )^{1/2} \,, \label{eom}
\end{equation}
where
\begin{mathletters}
\label{geexi}
\begin{equation}
{\cal G} = 1-\xi (s_1+s_2-2s_1s_2) \,, \label{calG}
\end{equation}
\begin{equation}
\xi = (2+\omega_{\rm BD})^{-1} \,.
\end{equation}
\end{mathletters}
For a circular orbit, the energy and orbital velocity are
given by
\begin{equation}
E=-(1/2) {\cal G} \mu m/r \,, \,\,
v^2 = {\cal G} m/r \,. \label{energy}
\end{equation}

\subsection{Gravitational radiation energy loss}

The rate of energy loss for a quasi-circular two-body orbit (that
is, circular, apart from an adiabatic inspiral) is given by
\begin{equation}
{{dE} \over {dt}}=-{8 \over 15}
{{\mu^2m^2} \over r^4} (12 \kappa v^2
+ {5 \over 8} \kappa_D {\cal S}^2 )\,, \label{edot}
\end{equation}
where the first term is the combined quad\-rupole/mono\-pole
contribution, and the second term is the dipole contribution,
and where
\begin{mathletters}
\label{parameters}
\begin{equation}
\kappa={\cal G}^2 (1- {1 \over 2} \xi
+ {1 \over 12} \xi \Gamma^2 ) \,, \label{kappa}
\end{equation}
\begin{equation}
\kappa_D=2 {\cal G}^2 \xi \,, \label{kappadee}
\end{equation}
\begin{equation}
{\cal S}=s_1-s_2 \,,
\end{equation}
\begin{equation}
\Gamma=1-2(m_1s_2+m_2s_1)/m \,.
\end{equation}
\end{mathletters}
The gravitational-waves are dominantly at the frequency $f=\omega/\pi$,
corresponding to twice the orbital frequency.

\subsection{Gravitational waveforms}

In the far-zone, the spatial components of the radiative metric
perturbation
${\bar h}^{\mu\nu} \equiv \eta^{\mu\nu}-\sqrt{(-g)} g^{\mu\nu}$
are given by
\begin{equation}
{\bar h}^{ij} = \theta^{ij} -{1 \over 2}
\theta \delta^{ij} - (\varphi/\phi_0)\delta^{ij} \,,
\end{equation}
where Greek and Roman indices denote spacetime and spatial components,
respectively, where $\varphi$ is the perturbation of the scalar
field $\phi$ about its asymptotic, cosmological value $\phi_0$
and where, to leading order in $v^2 \approx m/r$,
\begin{mathletters}
\label{thetaphi}
\begin{eqnarray}
\theta^{ij} = && 2(1-{1 \over 2}\xi)R^{-1} (d^2/dt^2)
\sum_A m_A x_A^ix_A^j \nonumber \\
= && (4\mu/R)(1-{1 \over 2} \xi) (v^iv^j-{\cal G}mx^ix^j/r^3) \,,
\end{eqnarray}
\begin{eqnarray}
\varphi/\phi_0  =&&  \xi (\mu/R)
\biggl\{ \Gamma [({\bf {\hat N} \cdot v})^2-{\cal G}m({\bf {\hat N}
\cdot x})^2/r^3 ]  \nonumber\\
&& -({\cal G}\Gamma +2\Lambda)m/r -2{\cal S}({\bf {\hat N}
    \cdot v}) \biggr\} \,,
\end{eqnarray}
\end{mathletters}
where $R$ and $\bf {\hat N}$ are the distance and direction unit
vector, respectively, of the observer, and $\Lambda=1-s_1-s_2+O(\xi)$.

The components of the Riemann tensor $R^{0i0j}$ measured by a
detector can be shown to be given by
$R^{0i0j}=-{1 \over 2} (d^2/dt^2) h^{ij}$, where $h^{ij}$
is the effective gravitational waveform, given by
\begin{equation}
h^{ij}=\theta_{TT}^{ij} - {1 \over 2} (\varphi/\phi_0)(\delta^{ij}
-{\hat N}^i{\hat N}^j) \,,
\end{equation}
where TT denotes the transverse-traceless projection.  Note that the
full gravitational waveform is transverse but not traceless because
of the presence of the scalar contribution.

For quasi-circular orbits, the waveform becomes
\begin{mathletters}
\label{waveform}
\begin{eqnarray}
h^{ij} =&& {{2\mu} \over R} \biggl[ Q_{TT}^{ij}
+ S(\delta^{ij}-{\hat N}^i{\hat N}^j) \biggr] \,, \label{waveform2}
\end{eqnarray}
\begin{eqnarray}
Q^{ij}= && 2(1- {1 \over 2} \xi){{{\cal G}m} \over r} ({\hat \lambda}^i
{\hat \lambda}^j-{\hat n}^i{\hat n}^j ) \,, \label{quad}
\end{eqnarray}
\begin{eqnarray}
S= &&-{1 \over 4}\xi \biggl\{ {{\Gamma {\cal G}m} \over r}
[(\bbox{ {\hat N} \cdot {\hat \lambda}})^2-(\bbox{ {\hat N}
\cdot {\hat n}})^2 ] -({\cal G}\Gamma +2\Lambda) {m \over r} \nonumber\\
&&-2{\cal S} ({{{\cal G}m} \over r})^{1/2} \bbox{ {\hat N} \cdot
{\hat \lambda}} \biggr\} \,, \label{scalar}
\end{eqnarray}
\end{mathletters}
where ${\bf {\hat n}} \equiv {\bf x}/r$,
and $\bbox{ {\hat \lambda}}\equiv{\bf v}/v$.

\section{Testing scalar-tensor gravity using matched filtering of
gravitational waveforms}

\subsection{Phase-shift estimate}

Because broad-band detectors such as the free-mass laser
interferometric systems detect the gravitational waveform $h^{ij}(t)$
superimposed on the noise, and because hundreds to tens of thousands
of cycles may be observed in the bandwidth, the observations are
especially sensitive to the evolution of the frequency and phase
of the wave.  By combining Eqs. (\ref{eom}), (\ref{energy})
and (\ref{edot}), one can show that the frequency of the waveform
evolves according to
\begin{equation}
\dot f={96 \over 5} \eta {{\cal G}^{1/2} \over {\pi m^2}}
\biggl({m \over r} \biggr)^{11/2}
\biggl( \kappa+{5 \over 96}{\kappa_D \over {\cal G}}
{r \over m} {\cal S}^2 \biggr) \,. \label{fdot}
\end{equation}
We define the Brans-Dicke chirp mass ${\cal M}$ and the dipole
parameter $b$ according to
\begin{eqnarray}
{\cal M} \equiv && (\kappa^{3/5} /{\cal G}^{4/5} )
\eta^{3/5} m \,, \nonumber \\
b \equiv &&(5/96)(\kappa^{-3/5} {\cal G}^{-6/5} )
\kappa_D {\cal S}^2 \,. \label{chirp}
\end{eqnarray}
Defining $u \equiv \pi {\cal M} f$, we put Eq. (\ref{fdot}) into
the form
\begin{equation}
\dot u = {\cal M}^{-1} (96/5) u^{11/3} (1+b \eta^{2/5} u^{-2/3}) \,.
\label{udot}
\end{equation}
Integrating, we get
\begin{equation}
u^{-8/3}[1-(4/5)b \eta^{2/5} u^{-2/3} ]=(256/5)(t_c-t)/{\cal M} \,,
\label{uintegral}
\end{equation}
where $t_c$ is the time at which $u \to \infty$.  We have expanded the
expression (\ref{uintegral})
to first order in $b \eta^{2/5} u^{-2/3}$, using the fact that
\begin{eqnarray}
b \eta^{2/5} u^{-2/3} \le
&& 5 \times 10^{-3} \, \biggl({500 \over \omega_{\rm BD}}\biggr)
   \biggl({{\cal S} \over 0.5}\biggr)^2 \nonumber \\
&& \times \biggl({M_\odot \over {\cal M}}\biggr)^{2/3}
   \biggl({{30 {\rm Hz}} \over f}\biggr)^{2/3} \,.
\end{eqnarray}
{}From Eq. (\ref{phase}), the number of cycles observed in a given
bandwidth can be written
$\Phi_{GW} = (2/{\cal M}) \int_{u_{\rm in}}^{u_{\rm out}} (u/{\dot u}) du$,
giving
\begin{equation}
\Phi_{GW}={1 \over 16} (u_{\rm in}^{-5/3}-u_{\rm out}^{-5/3})
-{5 \over 112} b\eta^{2/5}(u_{\rm in}^{-7/3}-u_{\rm out}^{-7/3}) \,.
\end{equation}
Demanding that the phase cont\-ri\-but\-ion of the di\-pole term be
no more than $\pi$, we obtain $b<(112\pi/5)\eta^{-2/5}u_{\rm in}^{7/3}$,
where we assume that $f_{\rm out}>> f_{\rm in}$ (1000~Hz vs.~10~Hz).
To lowest order in $1/\omega_{\rm BD}$, $\kappa={\cal G}=1$,
$\kappa_D=2/\omega_{\rm BD}$, and thus
$b \approx (5/48) {\cal S}^2/\omega_{\rm BD}$, resulting in the bound
given in Eq. (\ref{bound1}).

\subsection{Matched-filter analysis}

To obtain a more accurate estimate of the bound that can be placed on
the dipole parameter $b$, we carry out a full matched-filter analysis,
following the method described by Chernoff and Finn \cite{finn}
and Cutler and Flanagan \cite{cutler}.  To the accuracy needed,
we approximate the observed gravitational waveform, Eq. (\ref{waveform}),
in a given detector by $h(t) \approx {\cal R} \{ h^0(t) e^{i\Phi(t)} \}$,
where $h^0(t)$ is the slowly-varying Newtonian-order contribution to
the waveform amplitude, dependent upon the distance to the source,
its location on the sky, the orientation of the detector, and on
the source parameters $\cal M$, $\eta$ and $r$; $\Phi(t)$ is
the gravitational-wave phase, dominantly at twice the orbital
phase, and ${\cal R}$ denotes the real part.  The phase includes
the dipole and higher order post-Newtonian corrections, in principle.
Calculating the Fourier transform of $h(t)$ in the stationary phase
approximation, we obtain
\begin{equation}
{\tilde h}(f) = \cases{ {\cal A}f^{-7/6} e^{i\Psi} &$0<f<f_{max}$\cr
0&$f>f_{max}$ \,,\cr} \label{fourier}
\end{equation}
where ${\cal A} \propto R^{-1}{\cal M}^{5/6} \times$ [function of
angles and detector orientation], and $f_{max} \approx O(m^{-1})$
corresponds to the frequency when the inspiral turns into a plunge
toward coalescence, and
\begin{eqnarray}
\Psi(f)=&& 2\pi ft_c-\Phi_c-\pi/4 \nonumber \\
&&+{3 \over 128} u^{-5/3} (1-{4 \over
7}b\eta^{2/5}u^{-2/3} ) \,, \label{psi}
\end{eqnarray}
where $\Phi_c$ formally is the gravitational-wave phase at time $t_c$.

With a given noise spectrum $S_n(f)$, one defines the inner product of
signals $h_1$ and $h_2$, by
\begin{equation}
(h_1|h_2) \equiv 2 \int_0^\infty
{{{\tilde h}_1^* {\tilde h}_2+{\tilde h}_2^* {\tilde h}_1}
\over {S_n(f)}} df \,. \label{innerprod}
\end{equation}
The signal-to-noise ratio for a given signal $h$ is given by
\begin{equation}
\rho[h] \equiv S/N[h]=(h|h)^{1/2} \,. \label{sn}
\end{equation}
If the signal depends on a set of parameters $\theta^a$ which are to
be estimated by the matched filter, then the rms error in $\theta^a$
in the large $S/N$ limit is given by
\begin{equation}
\Delta\theta^a \equiv\sqrt {\langle (\Delta\theta^a)^2 \rangle}
= \sqrt{ \Sigma^{aa}} \,,
\end{equation}
where $\Sigma^{aa}$ is the corresponding component of the inverse
of the ``covariance matrix'' or ``Fisher information matrix'',
$\Gamma_{ab}$, defined by
\begin{equation}
\Gamma_{ab} \equiv \biggl( {{\partial h} \over {\partial \theta^a}}
\biggl\vert \biggr. {{\partial h} \over {\partial \theta^b}} \biggr) \,.
\label{fisher}
\end{equation}
The correlation coefficient between two parameters $\theta^a$ and
$\theta^b$ is
\begin{equation}
c^{ab}= \Sigma^{ab}/\sqrt{ \Sigma^{aa} \Sigma^{bb}} \,.
\label{correlation}
\end{equation}
For a noise spectrum, we adopt the analytic fit to the LIGO
``advanced detector'' noise spectral density,
\begin{equation}
S_n(f)= \cases{ \infty &$f<10$ Hz,\cr S_0[(f_0/f)^4+2(1+(f/f_0)^2)]
&$f>10$ Hz ,\cr} \label{noise}
\end{equation}
where $S_0=3 \times 10^{-48} \,{\rm Hz}^{-1}$ and $f_0=70\, {\rm Hz}$.
The cutoff at 10 Hz corresponds to seismic noise, while the $f^{-4}$
and $f^2$ dependences correspond to thermal and photon shot noise
respectively.

We shall adopt the following five parameters to be estimated:
$\ln {\cal A}$, $\Phi_c$, $f_0t_c$, $\ln {\cal M}$, and
$\tilde b$, where $\tilde b=b\eta^{2/5}$.  The corresponding partial
derivatives of $\tilde h (f)$ are
\begin{mathletters}
\label{partials}
\begin{eqnarray}
{{\partial \tilde h (f)} \over {\partial \ln {\cal A}}}
=&& \tilde h (f) \,,
\label{dhda}\\
{{\partial \tilde h (f)} \over {\partial f_0t_c}}
=&&  2\pi i (f/f_0) \tilde h (f) \,,
\label{dhdt}\\
{{\partial \tilde h (f)} \over {\partial \Phi_c}}
=&&  -i \tilde h (f) \,,
\label{dhdphi}\\
{{\partial \tilde h (f)} \over {\partial \ln {\cal M}}}
=&&  -{5 \over 128}i u^{-5/3} \tilde h (f)
[1- {4 \over 5} \tilde b u^{-2/3} ] \,,
\label{dhdm}\\
{{\partial \tilde h (f)} \over {\partial \tilde b }}
=&& -{3\over 224}i u^{-7/3} \tilde h (f) \,. \label{dhdb}
\end{eqnarray}
\end{mathletters}
The signal-to-noise ratio is given by
\begin{equation}
\rho^2= 4|{\cal A}|^2 f_0^{-4/3} I(7)/S_0 \,,
\end{equation}
where we define the integrals
\begin{equation}
I(q)  \equiv \int_{1/7}^\infty x^{-q/3} (x^{-4}+2+2x^2)^{-1} dx \,.
\end{equation}
We also define the coefficients $B_q \equiv I(q)/I(7)$.   Since our
{\it a priori} expectation is the validity of GR, we are looking for
a bound on $|\tilde b|$.  Equivalently, we
wish to determine the error in estimating $\tilde b$ about the nominal
value $\tilde b=0$, thus we set $\tilde b=0$ in Eq. (\ref{dhdm}).
Then the elements of the covariance matrix turn out to be proportional
to $\rho^2 u_0^{-n/3} B_q$ for various integers $n$ and $q$, where
$u_0=\pi {\cal M} f_0$.  Inverting the matrix to obtain $\Sigma^{ab}$,
we obtain from the elements $\Sigma^{{\cal M}{\cal M}}$, $\Sigma^{bb}$,
and $\Sigma^{{\cal M}b}$, $\Delta (\ln {\cal M}) = 57.6 u_0^{5/3}/\rho$,
$\Delta (\tilde b ) =53.2 u_0^{7/3}/\rho$, and
$c^{{\cal M} \tilde b} = -0.98$

Substituting $f_0=70 \,{\rm Hz}$, and working to first order in
$1/\omega_{\rm BD}$, so that ${\cal M}=\eta^{3/5}m$ and
$\tilde b=(10/96)\eta^{2/5} {\cal S}^2/\omega_{\rm BD}$
(see Eq. (\ref{chirp})), we obtain
\begin{mathletters}
\begin{equation}
\Delta(\ln {\cal M})= 6.56 \times 10^{-5} \biggl({{\cal M} \over
M_\odot}\biggr)^{5/3}
\biggl({10 \over {S/N}}\biggr) \,,
\end{equation}
\begin{equation}
\Delta \biggl({{\cal S}^2 \over \omega_{\rm BD}}\biggr)
= 6.14 \times 10^{-6} \eta^{-2/5}
\biggl({{\cal M} \over M_\odot}\biggr)^{7/3}
\biggl({10 \over {S/N}}\biggr) \,. \label{bound3}
\end{equation}
\end{mathletters}

\subsection{Post-Newtonian effects}

The interpretation of gravitational-wave observations as testing
scalar-tensor gravity relies upon determining the masses of the
two bodies with sufficient accuracy (i)~that one can decide with
confidence whether each body is more or less massive than
the accepted neutron-star maximum mass (modulo the many uncertainties
in that number) in the neutron-star/black-hole case, or (ii)~that
one can establish that the mass difference exceeds some critical
value, in the double neutron-star case.  The determination of
the individual masses makes use of post-Newtonian corrections
to the orbital phase, which depend explicitly on the reduced
mass parameter $\eta$.  It is then necessary to check whether such
determinations can be made {\it simultaneously} with the bound on
the parameter $\tilde b$.  To this end, we extend the matched filter
analysis to include post-Newtonian corrections.  Those corrections
have not been fully calculated to date in the context of BD, but
one expects them to be the same as those of GR, within corrections
of $O(1/\omega_{\rm BD})$.  To incorporate them into our analysis in
a first approximation, then, it suffices to add the appropriate
GR post-Newtonian terms of \cite{cutler}, Eq. (3.13), to the
Fourier transform phase, Eq. (\ref{psi}), to obtain
\begin{eqnarray}
\Psi(f) =&& 2\pi ft_c-\Phi_c-\pi/4 \nonumber \\
&&+{3 \over 128} u^{-5/3} \biggl[ 1-{4 \over
7}b\eta^{2/5}u^{-2/3} \nonumber \\
&& + {20 \over 9} ({743 \over 336}+{11 \over 4}
\eta) \eta^{-2/5} u^{2/3} \nonumber\\
&&-16\pi\eta^{-3/5}u \biggr] \,, \label{psipn}
\end{eqnarray}
where the final term is the ``tail'' effect.  Adding the parameter
$\ln \eta$ to the set to be estimated, we find the partial
derivatives
\begin{mathletters}
\begin{eqnarray}
{{\partial \tilde h (f)} \over {\partial \ln {\cal M}}}
=&&  -{{5i} \over 128} u^{-5/3} \tilde h (f)
\biggl[ 1- {4 \over 5} \tilde b u^{-2/3} \nonumber \\
&&+ {4 \over 3} ({743 \over 336}+{11 \over 4}
\eta) \eta^{-2/5} u^{2/3} \nonumber \\
&&-{32 \over 5}\pi\eta^{-3/5}u \biggr] \,, \label{dhdmnew}\\
{{\partial \tilde h (f)} \over {\partial \ln \eta}}
=&& {i \over 96} u^{-5/3} \tilde h (f)
\biggl[ (-{743 \over 168}+{33 \over 4}
\eta) \eta^{-2/5} u^{2/3} \nonumber \\
&&+{108 \over 5}\pi\eta^{-3/5}u \biggr] \,.
\label{dhdeta}
\end{eqnarray}
\end{mathletters}
The other partial derivatives in Eqs. (\ref{partials}) are unchanged.
Setting $\tilde b=0$, we then calculate and invert the covariance
matrix and evaluate the errors in the five relevant parameters
(the parameter $\ln {\cal A}$ decouples from the rest, and is not
important for our purposes), along with the correlation coefficients
between $\cal M$, $\eta$ and $\tilde b$.  Notice that the fact that
$\tilde b$ involves $\eta$ implicitly does not affect the outcome,
because we are considering the results centered around $\tilde b=0$.
For various double neutron-star and neutron-star/black-hole
systems, the results are shown in Table \ref{table1}.
Note that $\tilde b$ is less strongly correlated with
$\eta$ than is $\cal M$,
a result to be expected because of the very different
dependences of the dipole and post-Newtonian effects
on $m/r$ or on $u$.  Nevertheless, a result of the correlation is to
weaken
the bound
on ${\cal S}^2 /\omega_{\rm BD}$ in Eq. (\ref{bound3}) by a factor of about
2.3.
Eq. (\ref{fit}) provides an analytic fit to the result.

Notice from Table \ref{table1} that the accuracy in determining
$\eta$ is a few percent, consistent with the results of \cite{cutler}.
Together with the high
accuracy in determining $\cal M$, this will lead to accurate values
for the two masses, except for the degenerate region around equal
masses, $\eta=0.25$.  However, that region is excluded from our
considerations because of the weakness of the bound in the
nearly-equal-mass, double neutron-star case, and because of the
impossibility of determining unambiguously that one of two objects
of nearly equal mass is a black hole while the other is not,
in the mixed case.  Figure \ref{fig3} shows the regions in
${\cal M}-\eta$-space corresponding to the three types of system.

\subsection{Dependence on the nature of the system}

The nature of the system being observed is important to the
interpretation of the bound (\ref{fit}) as a test of gravitation
theory.

For two neutron stars, the chirp mass is relatively small
($1.2 M_\odot$ for two $1.4 M_\odot$ neutron stars), so the bound
on ${\cal S}^2/\omega_{\rm BD}$ can be quite small.  The reason is that
such systems enter the detectors' bandwidth (say, at 10 Hz) with a
large separation relative to the total mass ($r/m \approx 180$ for
two $1.4 M_\odot$ neutron stars).  As a result, gravitational radiation
damping is weaker, and more cycles occur in the detectors' bandwidth
(up to 16,500 cycles for two $1.4 M_\odot$ neutron stars), hence the
matched filter is more sensitive to effects on the template phase
evolution.  However, the sensitivity difference $\cal S$ is also
small, because of the weak dependence of the sensitivity on the
masses of neutron stars, and because the neutron stars are expected
to have similar masses, as has been seen in known binary pulsar
systems, such as the Hulse-Taylor system PSR 1913+16 \cite{note1}.
For relatively stiff equations of state, which are required in order
to have neutron stars sufficiently massive ($\ge 1.4 M_\odot$)
to agree with masses inferred from known binary pulsars, Table 3
of \cite{zaglauer} shows that the sensitivity varies roughly linearly
with mass, with $s \sim 0.2 m/M_\odot$, reaching a maximum near the maximum
mass for the given equation of state (around $1.46 M_\odot$
for the stiff equations of state discussed in \cite{zaglauer}).
To get a rough idea of the bounds on $\omega_{\rm BD}$ that might be
possible in double neutron-star systems, we substitute the relation
${\cal S} \approx 0.2 \beta \delta m/M_\odot$ into Eq. (\ref{fit}),
where $\delta m \equiv m_1-m_2$, and where $\beta$ may vary
between 0 and 1, and use the fact that $\delta m^2 =(1-4\eta)m^2$,
to obtain
\begin{equation}
\omega_{\rm BD} > 2736 \beta^2 \biggl({ M_\odot \over m}\biggr)^{1/3}
\biggl({1 \over \eta}-4 \biggr) \biggl({{S/N} \over 10} \biggr) \,.
\end{equation}
The resulting bounds for $\beta \approx 1$ are plotted in
Fig. \ref{fig2}.  For $\beta \approx 1$, the bounds are weaker
than solar-system bounds for neutron stars whose masses differ
by less than $0.5 M_\odot$.  At the other extreme, the bound
could approach 1000 for a $0.7 - 1.4 M_\odot$ pair.

For black holes in scalar-tensor gravity, $s_{BH}\equiv 0.5$.
This is a consequence of the fact that, in the formation of a
black hole, the scalar field is radiated away
(see \cite{zaglauer,hawking,dykla,zaglauer2} for further
discussion).  As a result, from Eqs. (\ref{geexi}) and
(\ref{parameters}), we see that  ${\cal G} =1-\xi/2$,
$\kappa=(1-\xi/2)^3$, ${\cal S}=\Gamma=\Lambda=0$.  Substituting
these values into Eqs. (\ref{eom}), (\ref{energy}), (\ref{edot})
and (\ref{thetaphi}), we see that the dipole effects vanish, and
the equations become equivalent to those of GR if we replace every
mass by the ``tensor mass'' \cite{hawking}, given by
\begin{equation}
m_T \equiv m(1-\xi/2)=m(3+2\omega_{\rm BD})/(4+2\omega_{\rm BD}) \,.
\end{equation}
If gravitational radiation is the only information about the system,
then only the tensor masses are measured in a matched filter.  Since
the behavior of the system is general relativistic in terms of tensor
masses, then no test of BD in double black-hole systems is possible.

For the case of neutron\--star/\-black\--hole sys\-tems, $s_{BH} =0.5$
while $s_{NS} \approx 0.1 - 0.2$.  Although the black hole cannot
support scalar field in its vicinity, the companion neutron star can,
and consequently dipole gravitational radiation can occur.  For stiff
equations of state, $s_{NS} \le 0.14$, and thus we can approximate
${\cal S} \ge 0.3$.  With these assumptions, Eq. (\ref{bound2}) leads
to the bounds plotted in Fig. \ref{fig1}.

\section{Testing scalar-tensor gravity using waveform amplitude and
polarization}

{}From Sec. II.C we see that scalar-tensor gravity affects the
amplitude of the gravitational waveform in two important ways:
(i)~it introduces a third, non-traceless polarization state
(Eq. (\ref{waveform2})), and (ii) it introduces, through the
dipole term, a contribution at the orbital frequency in addition
to the quadrupole/monopole contributions at twice the orbital
frequency (final term in Eq. (\ref{scalar})).   However, these
are unlikely to be testable, for the following reason.
{}From Eqs. (\ref{innerprod}) and (\ref{partials}), it is simple
to see that the cross-components between $\ln {\cal A}$ and the
other four parameters in the covariance matrix vanish, so that
$\Gamma_{ab}$ has the form
\begin{equation}
\Gamma_{ab}= \rho^2 \pmatrix{1&0\cr 0&D\cr} \,,
\end{equation}
where $D$ is a $4 \times 4$ matrix corresponding to the parameters
$\Phi_c$, $f_0t_c$, $\ln {\cal M}$, and $\tilde b$.  Thus
$\Delta (\ln {\cal A})= (\Sigma^{{\cal A}{\cal A}})^{1/2}
=1/\rho\approx 0.1$, for a signal-to-noise ratio of 10.
However, the scalar-tensor corrections to the amplitude are
much smaller than this.  The ratio of the monopole part of
$S$ (Eq.  (\ref{scalar})) to the quadrupole term $Q^{ij}$
(Eq. (\ref{quad})) is
$O(\xi/8) \approx 2\times 10^{-4} (500/\omega_{\rm BD})$,
while the ratio of the dipole part of $S$ to the quadrupole term
is $O[{1 \over 4} \xi {\cal S} (r/m)^{1/2}] \approx 10^{-3}
(500/\omega_{\rm BD})({\cal S}/0.3)(r/100m)^{1/2}$.
Thus, despite the presence of qualitatively new contributions
to the waveform, the lower sensitivity of matched filtering
to amplitudes, as compared to phases, makes those contributions
difficult to detect or bound.

\section{Discussion of results}

We have shown that interesting, if not spectacular bounds on the
coupling constant $\omega_{\rm BD}$ of Brans-Dicke theory could be
obtained by matched filtering of a gravitational wave signal
from an inspiraling binary of a neutron star and either a neutron
star of very different mass, or a low-mass black hole.
The bound that can be obtained decreases rapidly with decreasing
neutron-star mass difference and with increasing black-hole mass.

Another class of laser-interferometric detectors that may be relevant
to this discussion is the space-based class of systems, such as LAGOS
\cite{lagos} and SAGITTARIUS/LISAG \cite{sagittarius}.
These will be most sensitive to
gravitational waves in the frequency range $10^{-4}$ to $10^{-1}$ Hz,
and could detect gravitational waves from the inspiral of a $1
M_\odot$ star into a black hole in the $10^6$ -- $10^7 \, M_\odot$
range.  Noting that the bound on $\omega_{\rm BD}$ scales with the
frequency $f_0$ as $f_0^{-7/3}$, we find from Eq. (\ref{bound2}) for a
$1 M_\odot$ star and a $10^6 M_\odot$ black hole, with $f_0 \approx
10^{-2}$ Hz, that a bound exceeding 50,000 might be possible.  This
warrants a more detailed matched-filter study, including
the use
of the appropriate noise
curve for space-based detectors analogous to Eq. (\ref{noise}), and
including the effects of spin of the central hole, which are likely to
be important.  This study is currently under way.

It is useful to point out that, for generalized scalar-tensor (ST)
theories, {\it i.e.} those in which the coupling constant becomes
effectively a function of the scalar field \cite{wagoner},
the foregoing conclusions still apply, with the following principal
change: the dipole parameter $\kappa_D$, Eq. (\ref{kappadee}),
becomes
\begin{equation}
\kappa_D={{2{\cal G}^2} \over {(2+\omega_{\rm ST})}}
\biggl( 1+{{2\omega_{\rm ST}^\prime}
\over {(3+2\omega_{\rm ST})^2}} \biggr)^2 \,,
\end{equation}
where $\omega_{\rm ST}^\prime \equiv d\omega_{\rm ST} /d\phi$.
(There are other changes induced by $\omega_{\rm ST}^\prime$,
in $\kappa$, $\Gamma$ and other formulae, but they are
unimportant for our purposes.)  In the large-$\omega_{\rm ST}$
limit, this means replacing $\omega_{\rm BD}$
by $\omega_{\rm ST}/(1+\omega_{\rm ST}^\prime/2\omega_{\rm ST}^2)^2$ in the
bounds
of Eqs. (\ref{bound1}), (\ref{bound2}) and (\ref{bound3}).

Another difference between BD and such generalized scalar-tensor
theories is that, while in the former case, large $\omega_{\rm BD}$
implies that all physical predictions are close to those of GR
(roughly within $O(1/\omega_{\rm BD})$), in the latter case the statement
holds only for weak scalar-field situations ($G\phi \approx 1$),
such as in the solar system.  However, in the strong-field interiors
of neutron stars, or in the early universe, where the scalar field
may have values very different from its weak-field values, the
differences between the scalar-tensor theory and GR may be
significant, despite a large exterior value of
$\omega_{\rm ST}$ \cite{zaglauer2,esposito,damtaylor,esposito2}.
In evaluating the sensitivities of neutron stars, we used neutron
star models computed using GR, since BD is a small correction
for large $\omega_{\rm BD}$ throughout the stellar interior \cite{salmona}.
But the strong-field effects in ST theories alter the situation,
and quite different values of sensitivities for a given neutron
star mass may result \cite{zaglauer2,esposito2}.  Since our quoted
bounds were strongly dependent on the sensitivities, it is impossible
to draw strong conclusions about bounds on ST theories from
inspiralling binaries without further research.

How do the bounds we have suggested compare with other empirical
bounds and with theoretical expectations?   Solar-system bounds
on scalar-tensor gravity may improve from the current level of
$\omega_{\rm BD}>500$ during the coming decade.  The Relativity
Gyroscope Experiment (Gravity Probe B) aims to measure
the geodetic precession of a set of orbiting superconducting
gyroscopes at the $10^{-5}$ level, which would correspond to
$\omega_{\rm BD}>60,000$ \cite{everitt}.  Orbiting optical
interferometers for astrometry may measure the deflection
of light at the 5 microarcsecond level, resulting in bounds
on $\omega_{\rm BD}$ at the level of $10^5$ \cite{reasenberg}.

In BD, $\omega_{\rm BD}$ is an arbitrary constant, so no value has
an {\it a priori} theoretical significance (except
$\omega_{\rm BD}=\infty$).  However, in generalized theories,
the value of $\omega_{\rm ST}$ is coupled to the dynamics of the
scalar field (or fields, in multi-scalar theories \cite{esposito}),
and its present, average cosmic value may depend on the evolution
of the universe.  Damour and Nordtvedt \cite{damnordt} have pointed
out that, in one class of generalized scalar-tensor theories, in which
$\omega(\phi) \equiv -{3 \over 2} - {1 \over 2} (\chi \ln \phi)^{-1}$,
with $0 \le \phi \le 1$, cosmic evolution from strongly
non-general-relativistic early universes tends toward a
large-$\omega$ ``attractor'' at the present epoch.
For specific models and values of the parameter $\chi$,
the present value of $\omega_{\rm ST}$ can range from
$2 \times 10^4$ to $10^6$.  Unless there are unusual strong-field
effects in this theory that would modify our conclusions based
on BD, these values are well above the bounds we anticipate from
coalescing binaries.

\acknowledgments

I am grateful to Kip Thorne and Curt Cutler for especially useful
comments.
This work is supported in part by the National Science Foundation
under Grant No. 92-22902, and the National Aeronautics and Space
Administration under Grant No. NAGW 3874.

\begin{figure}
\caption{Bounds on $\omega_{\rm BD}$ from inspir\-al\-ling
neu\-tron\--star/\-black-hole binaries, plotted against black-hole mass,
for various neutron-star masses.  Hatched portion indicates black holes
with mass less than $3.0 M_\odot$, where identification as a
black-hole may be ambiguous.  Curves assume ${\cal S}=0.3$ and a
signal\--to\--noise ratio of 10.}
\label{fig1}
\end{figure}

\begin{figure}
\caption{Bounds on $\omega_{\rm BD}$ from inspir\-al\-ling double
neu\-tron\--star binaries, plotted against mass of neutron stars.
For equal masses, dipole radiation is suppressed, and no bound
on $\omega_{\rm BD}$ results.  Curves assume a linear dependence of
sensitivity on mass, with ${\cal S}=0.2 \delta m /M_\odot$, and a
signal-to-noise ratio of 10.}
\label{fig2}
\end{figure}

\begin{figure}
\caption{Chirp-mass/reduced\--mass param\-eter plane, showing location
of three different types of compact binaries.  Chirp mass is plotted in
units of the neutron-star maximum mass.}
\label{fig3}
\end{figure}

\twocolumn[
\widetext
\begin{table}
\caption{The rms errors for signal parameters, the corresponding
bound on $\omega_{\rm BD}$, and the correlation coefficients
$c_{{\cal M}\eta}$, $c_{{\cal M} \tilde b}$ and $c_{ \tilde b \eta}$.
General relativistic post-Newtonian effects are included; the noise
spectrum is that of the advanced LIGO system, given by
Eq. (\protect \ref{noise}), and a signal-to-noise ratio of 10
is assumed.  Masses are in units of $M_\odot$, $\Delta t_c$ is
in msec.  For neutron-star/black hole systems, ${\cal S}=0.3$
is assumed; while for double neutron-star systems,
${\cal S}=0.2 \delta m/M_\odot$ is assumed.}
\begin{tabular}{cccccccccc}
$m_1$&$m_2$&$\Delta \phi_c$&$\Delta t_c$&$\Delta {\cal M}/{\cal M}$
&$\Delta \eta/\eta$&$\omega_{\rm BD}$&$c_{{\cal M}\eta}$&$c_{{\cal M}
\tilde b}$&$c_{\tilde b \eta}$\\
\hline
\multicolumn{10}{c}{Neutron-star/black-hole systems}\\
1.0&2.0&2.21&0.83&0.028\%&2.23\%&2140&-0.950&-0.991&-0.909\\
1.0&5.0&2.21&0.83&0.056\%&2.20\%&682&-0.949&0.991&0.908\\
1.4&5.0&2.19&0.82&0.079\%&2.73\%&470&-0.953&-0.991&-0.911\\
1.4&10.0&2.19&0.82&0.131\%&2.46\%&194&-0.952&0.991&0.909\\
\hline
\multicolumn{10}{c}{Double neutron-star systems}\\
0.7&1.4&2.24&0.83&0.015\%&1.84\%&1092&-0.945&0.992&-0.905\\
1.4&1.4&2.19&0.82&0.029\%&2.34\%&0&-0.953&-0.991&-0.911\\
\end{tabular}
\label{table1}
\end{table}
]

\begin{references}
\bibitem{note1}
This suppression of dipole
radiation because of the similarity of the neutron stars is the reason
that the binary pulsar PSR 1913+16, despite the high timing accuracy,
gives a bound on $\omega_{\rm BD}$ that is weaker than the solar-system
bound \cite{zaglauer}.  In that case, the bound arises from the
combination of the periastron shift and the quadrupole-monopole
parts of the energy loss.  The periastron shift will not be relevant
for inspiralling binaries because the orbits are expected to be
circularized long before the radiation enters the detectors'
bandwidth.
\end{references}
\end{document}